# Hybrid GA/MSDLO Approach to Solve a Greenfield Robotized Machine Tending Layout

Ramez Awad[a,]*, Joshua Beck[a]

[a]Fraunhofer IPA, Nobelstr. 12, 70569 Stuttgart, Germany
* Ramez Awad. Tel.: +49-711-970-1844. *E-mail address:* ramez.awad@ipa.fraunhofer.de

**Abstract**

Meta-heuristics like genetic algorithms (GAs) effectively explore solution spaces but often yield suboptimal results and are sensitive to parameter tuning. In contrast, the mass-spring-damper layout optimization (MSDLO) method is fast and excels in locally optimizing resource arrangements, though its effectiveness depends on the initial resource placement. This paper first employs genetic algorithms to establish an optimal resource placement in a greenfield layout. We then apply MSDLO to fine-tune the positions and orientations of these resources. The results are presented and discussed, highlighting this approach's practicality in effectively addressing industry needs.

## 1. Introduction

Voigt describes the planning and optimization of factory layout as "the optimized decision concerning the type, quantity, and spatial arrangement of production resources" (translated to English from [2]). A well-designed layout enhances workflow, minimizes handling costs, and boosts productivity, while a poorly designed layout results in higher defects, waste, and increased costs [3–5]. The challenge in creating an effective layout is that the solution space for this decision problem is usually continuous and vast, complicating the resolution process [6, 7].

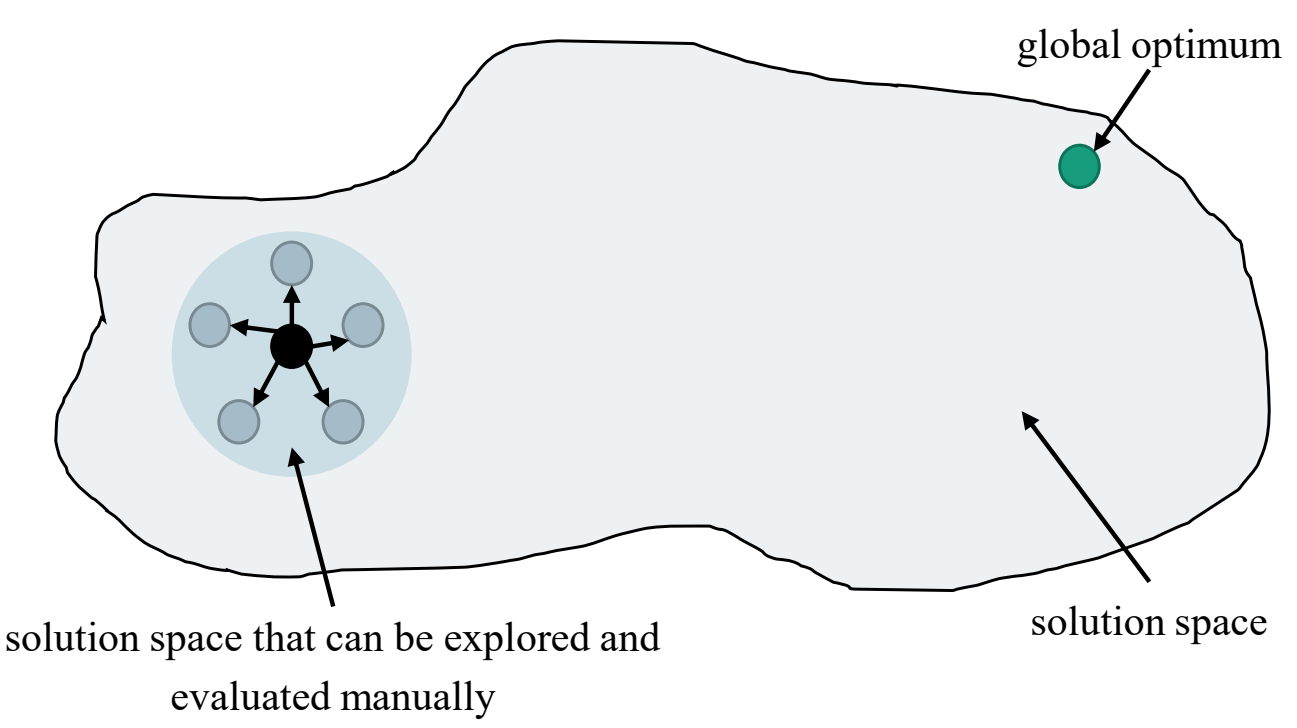

Fig. 1. Solution space explorable by production engineer compared to overall solution space (modelled after [1])

Accordingly, production engineers often rely on traditional patterns when creating new layouts rather than exploring the entire solution space for the best possible solution [1], see Fig. 1. This results in operating costs wasted. Therefore, a method is needed to systematically explore a large portion of the solution space within a limited timeframe, either by comparing solution alternatives or directly identifying the optimal solution.

[8] presented the Mass-Spring-Damper Layout Optimization (MSDLO) method, which converts the process interdependencies between resources and the financial considerations of a layout problem into a Mass-Spring-Damper system. The equilibrium of this system is then translated back into the arrangement of resources within a layout solution.

[8] evaluated the performance of the MSDLO method against a genetic algorithm (GA) for a small Greenfield layout problem (n = 5 resources) using two key performance indicators: (1) Handling Cost, which refers to the expense of moving parts to and from resources during a planning period, and (2) Area Savings, representing "fictitious" revenues based on the premise that unused space can be "rented" to other departments within the company. A quadratic savings function emphasizes the flexible usage of larger areas.

The evaluation revealed that the MSDLO method excelled in optimizing Handling Cost, while the GA outperformed in optimizing Area Savings. Additionally, it was found that the performance of the MSDLO method is highly dependent on the initial placement of the resources.

In this paper, the approaches are combined sequentially. The GA serves as a coarse optimizer to achieve an advantageous initial placement of resources. Subsequently, the MSDLO method is employed as a local planner to refine the layout solution. This hybrid approach is evaluated against the same Greenfield layout problem used in [8].

The paper is structured as follows: In section 2 the state-of-the-art regarding layout optimization is summarized. Section 3 recaps the Greenfield layout problem. The GA/MSDLO solution approach is outlined in section 4. The results of the approach are presented and discussed in section 5. Conclusions and future work are listed in section 6.

## 2. Related Work

Layout planning and optimization within the context of robotized production lines is a specific aspect of the general "facility layout problem" (FLP). The FLP has been established as NP-Hard, as noted in various studies [9–11]. A detailed overview of the various aspects and solution approaches of the FLP can be found in [10] and [12]. The solution approaches can be classified into three categories:

1) Exact methods seek to find the global optimal solution to the FLP [9]. The literature identifies three main algorithm classes:
   - Cutting Plane Algorithm (CPA)
   - Branch & Bound Algorithm (B&B)
   - Dynamic Programming (DP)
2) Heuristic approaches which aim to build up the layout by iteratively allocating machines or departments, or to improve upon a given layout solution [13–15].
3) Meta-heuristics, which are “high-level problem-independent algorithmic framework[s] that [provide] a set of guidelines or strategies to develop heuristic optimization algorithms” [16]. Notable meta-heuristics include:
   - Genetic Algorithms (GAs)
   - Simulated Annealing (SA)
   - Tabu Search (TS)
   - Ant Colony Optimization (ACO)
   - Particle Swarm Optimization (PSO)

   Note that hybrid approaches, which combine meta-heuristics, exist.

The following will concentrate on the named exact methods and meta-heuristics.

***Cutting Plane Algorithms (CPA)***

The Cutting Plane Algorithm, initially proposed by Kelley for "convex programs," is utilized to address mixed integer problems. The process involves relaxing integer constraints, checking solutions, generating cuts to eliminate non-integer solutions, and iterating until an integer solution is reached [17, 18]. It shows an exponential increase in average runtime in solving problems involving 15+ facilities [19, 20].

***Branch & Bound Algorithm (B&B)***

Introduced by Land and Doig in 1960, the B&B algorithm divides the problem into subproblems, calculates bounds for each, and prunes branches that cannot yield better solutions [21, 22]. Variants of B&B have been successfully applied to facility layout problems, demonstrating their effectiveness for smaller instances (up to 15 departments; even 36 departments if symmetry constraints can be applied to reduce complexity). However, as the number of departments increases, computational feasibility diminishes, necessitating the use of heuristics for practical applications [23].

***Dynamic Programming (DP)***

Dynamic Programming, introduced by Richard Bellman, breaks complex problems into simpler overlapping subproblems, storing results to avoid redundancy, and finally constructs the final solution once the main problem is solved [24]. Though it is suitable for dynamic facility layout problems, DP often integrates heuristic methods to manage extensive state spaces, which may compromise the guarantee of global optimality [25].

***Genetic Algorithms (GAs)***

Inspired by natural selection, GAs operate through a cycle of population initialization, evaluation, selection, crossover, mutation, and replacement [26]. They effectively explore large solution spaces, often yielding near-optimal layouts [27, 28]. However, computational intensity and premature convergence remain challenges [28–30].

***Simulated Annealing (SA)***

SA mimics the annealing process in metallurgy, allowing solutions to escape local optima [31]. While SA has shown success in various layout optimization applications [32, 33], it can be slow to converge and sensitive to parameter choices [34].

***Tabu Search (TS)***

TS enhances local search algorithms to explore beyond local optima through a tabu list that prevents cycling back to previously explored solutions [35]. Its effectiveness has been demonstrated in various facility layout problems, although performance can be sensitive to parameter settings [36].

***Ant Colony Optimization (ACO)***

ACO leverages the foraging behavior of ants to find optimal solutions by exploring paths based on pheromone levels and heuristic information [37]. It has been successfully applied to various layout problems, but struggles with parameter sensitivity and scalability [38, 39].

***Particle Swarm Optimization (PSO)***

Developed based on social behavior patterns, PSO leverages the collaborative behavior of a swarm to explore the solution space efficiently [40]. While it performs well for smaller problems, it often faces challenges with larger instances, including longer computation times and local optima entrapment [41].

Table 1 summarizes how well the various state-of-the-art perform. This overview indicates that while exact methods are suitable for smaller problems, meta-heuristics are preferable for larger, more complex layouts, balancing optimality and computational efficiency.

Table 1. Summary of methods for layout optimization.

| Method | Type | Global Optimality | Scalability | Computational Efficiency |
|---|---|---|---|---|
| CPA | Exact | Yes | Limited | Poor |
| B&B | Exact | Yes | Limited | Poor |
| DP | Exact | Yes | Limited | Poor |
| GA | Meta-Heuristic | Near-optimal | Moderate | High |
| SA | Meta-Heuristic | Near-optimal | Moderate | Moderate |
| TSO | Meta-Heuristic | Near-optimal | Good | Moderate |
| ACO | Meta-Heuristic | Near-optimal | Good | Moderate |
| PSO | Meta-Heuristic | Near-optimal | Moderate | High |

## 3. Greenfield Layout Problem

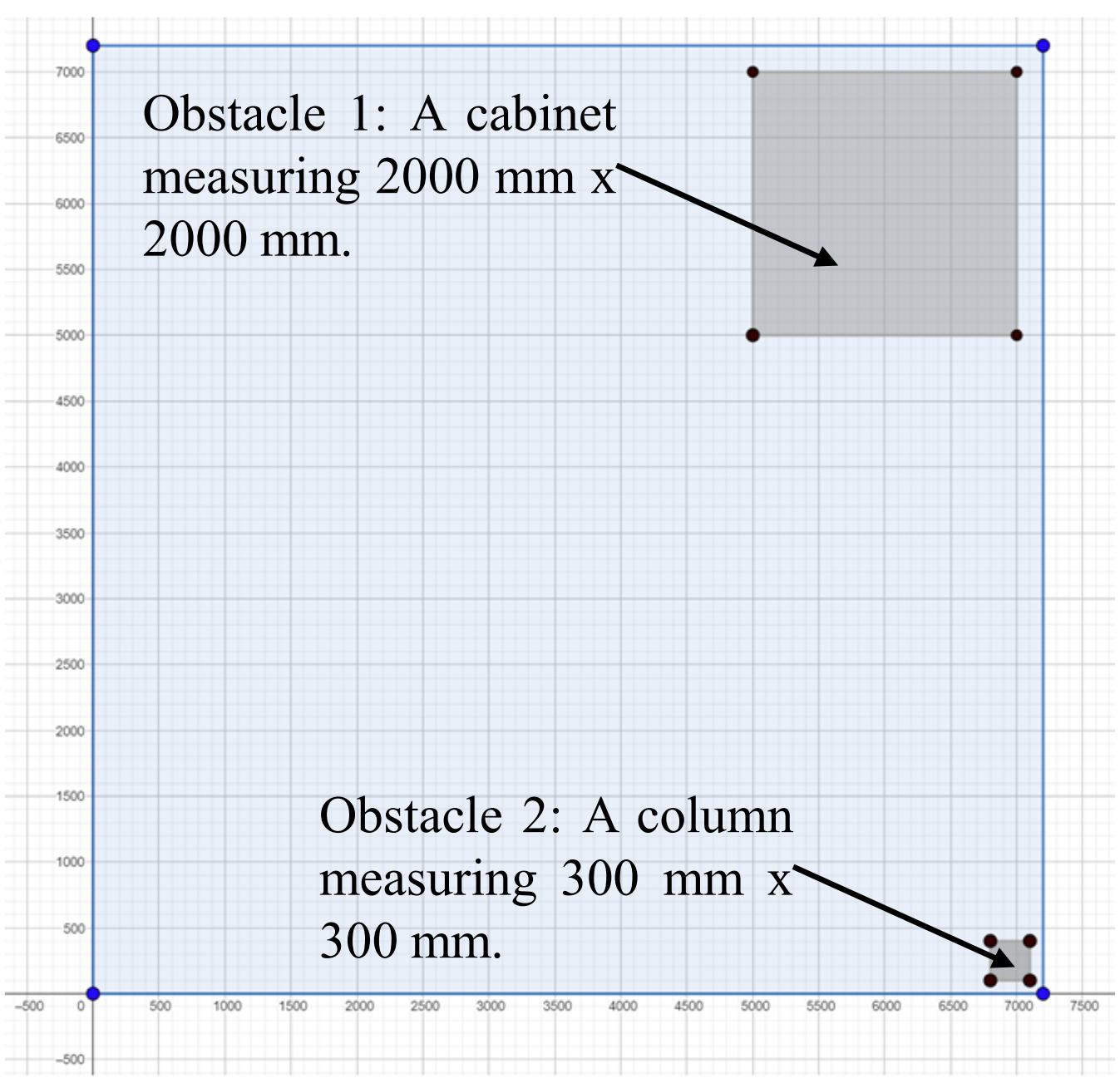

Fig.2. Greefield layout is a 7200 mm x 7200 mm square, featuring two pre-existing obstacles.

The layout problem used to evaluate the hybrid approach is displayed in Fig. 2.

The process to be carried out utilizes two CNC machines, one inspection machine, one material rack, and one robot with an attached gripper and an integrated electrical cabinet at its base (see Fig. 3).

The process consists of four handling processes:

- Transport part from the Material Rack (raw material shelf) to the CNC Machine 1.
- Transport part from THE CNC Machine 1 to THE CNC Machine 2.
- Transport part from the CNC Machine 2 to the Inspection Machine.
- Transport part from the Inspection Machine to the Material Rack (either into the shelf for “OK” parts or to the shelf for “NOK” parts.

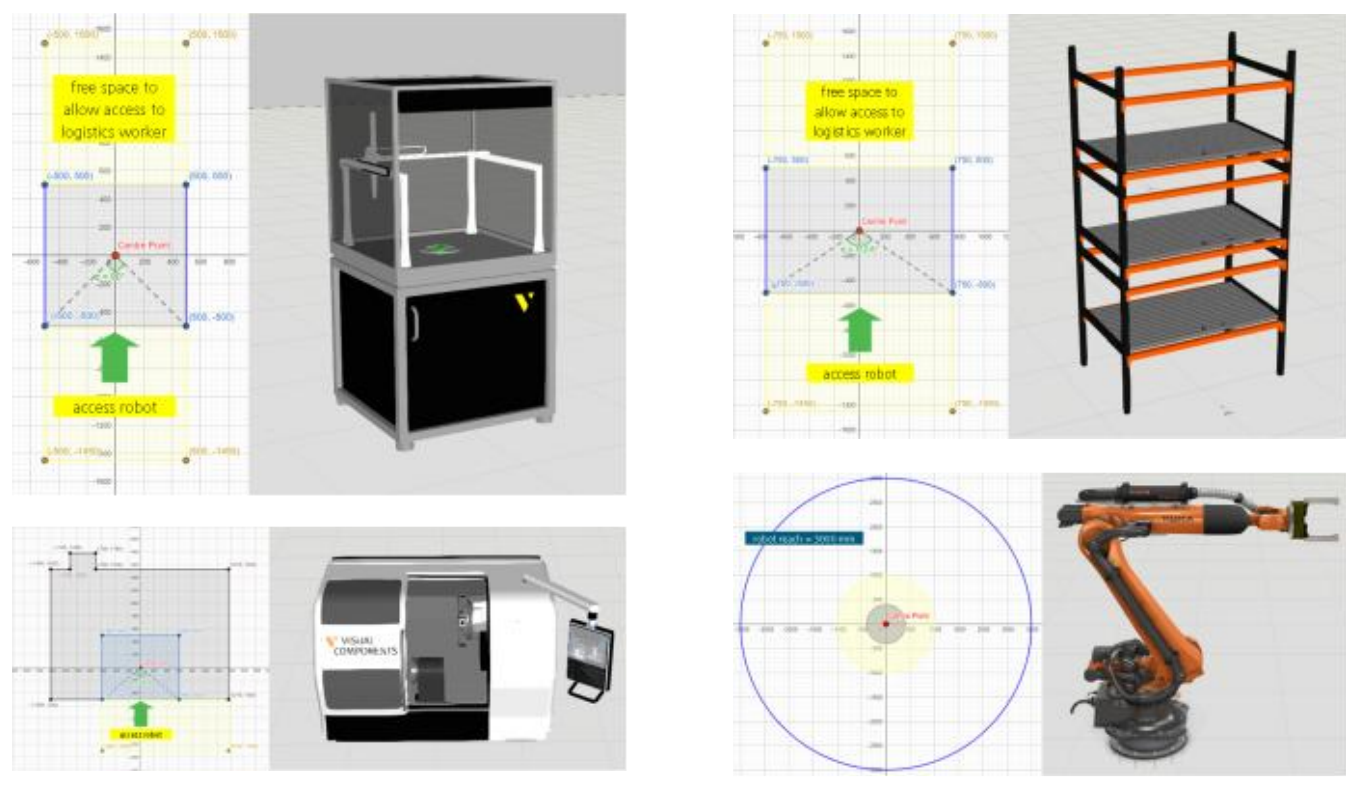

Fig. 3. 2D specifications and 3D models of the resources to be arranged in the layout.

The planning horizon is for 12 x 4 standard work weeks, with 8-hour shifts. The production volume for the planning period is 68,640 units. The two KPIs used for evaluation are:

**Handling Cost ($C_H$):**

$$C_H = \sum_v \sum_i \sum_j vol_v * itr_{ijv} * d_{ij} * c \qquad (1)$$

- $itr_{ijv}$ [#]: The required number of handling processes between the source resource ***i*** and destination resource ***j*** to produce a unit of a product variant ***v***.
- $vol_v$ [#]: The production volume of the product variant ***v*** in a planning period.
- $d_{ij}$ [m]: The route distance between the source resource ***i*** and destination resource ***j***.
- $c$ [€/m]: Average cost factor for moving a robot (specifically -1.25 €/km).

**Area Savings ($A_S$):**

$$A_S = \sum_a A_a * A_a * c_A \qquad (2)$$

- $A_a$[m²]: Usable area ***a***. For an area to be useful, it must have a minimum size of 1.00 m², at least 1.00 m of continuous shared boundary with the layout’s boundary and be at least 0.50 m wide.
- $c_A$ [€/m⁴]: The area savings factor (specifically +0.67 €/m⁴).

They are summed up in the **Aggregated KPI**:

$$Agrregated\ KPI = C_H + A_S \qquad (3)$$

## 4. The GA/MSDLO Hybrid Approach

In a first step, the Greenfield layout problem is solved via a GA.

Next, GA's solution is fed into the MSDLO pipeline (see Fig. 4).

- The layout's boundaries and obstacles are modelled as unmovable collision bodies (mass = infinity).
- The resources are modelled as collision bodies and placed in the layout using the GA's results.
- "Reachability" Spring-Damper-Links are added between the handling and non-handling resources. The spring constant is configured based on the number of handling process iterations for the planning horizon.
- "Closeness" Spring-Damper-Links are added between two non-handling resources being serviced by the same handling resource after each other. The spring constant is configured based on the number of handling process iterations for the planning horizon, multiplied by a closeness factor to account for the fact that those springs optimize the radius of the rotational movement of the handling resource.

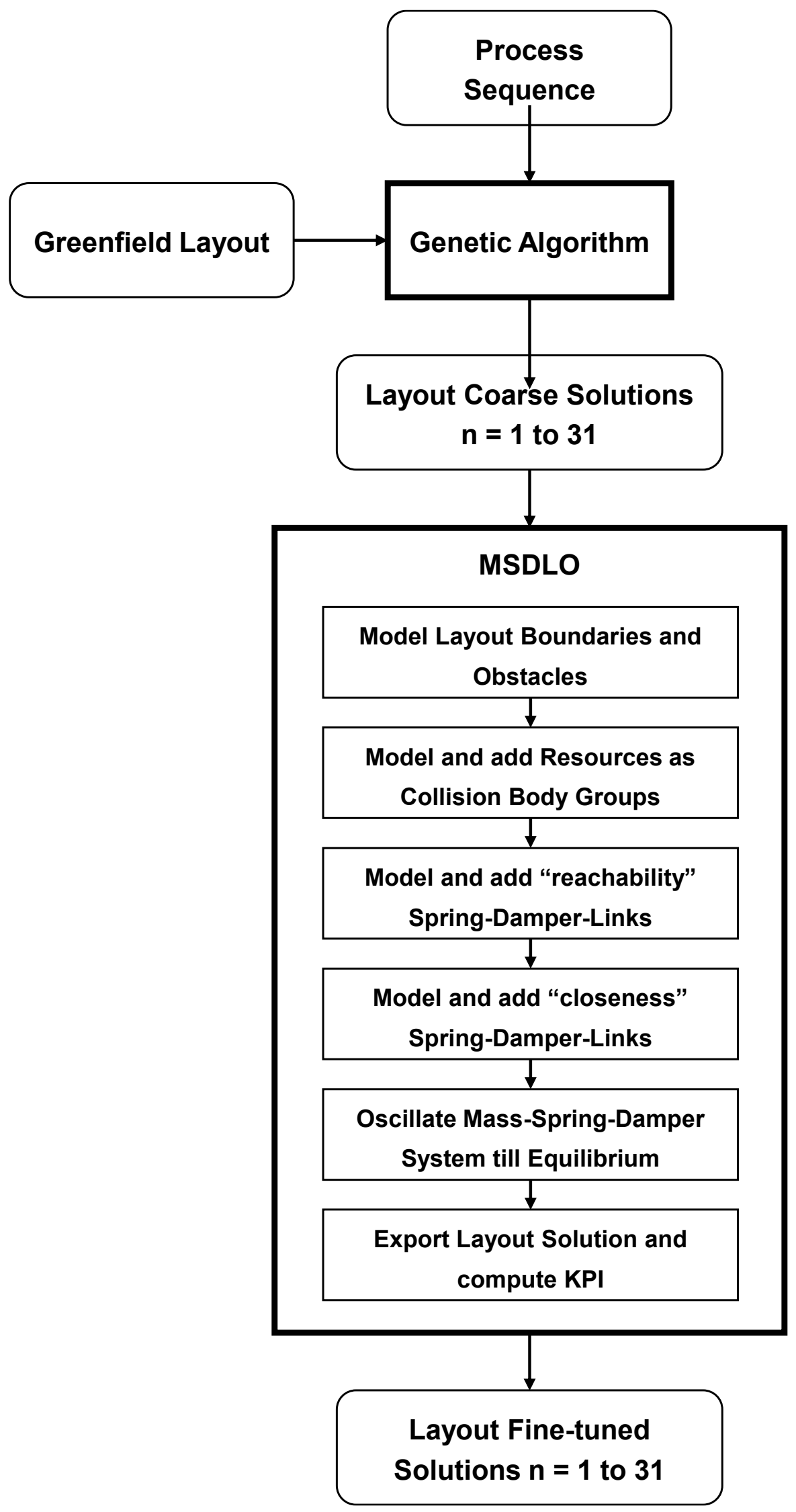

Fig. 4. Block diagram of the GA/MSDLO Hybrid Approach

- Once all collision bodies and spring-damper-links are modelled (see Fig. 5), the system is left to oscillate until it reaches its equilibrium.
- The position and rotation of the resources in the equilibrium are extracted, and the KPIs and the aggregated KPI are computed.

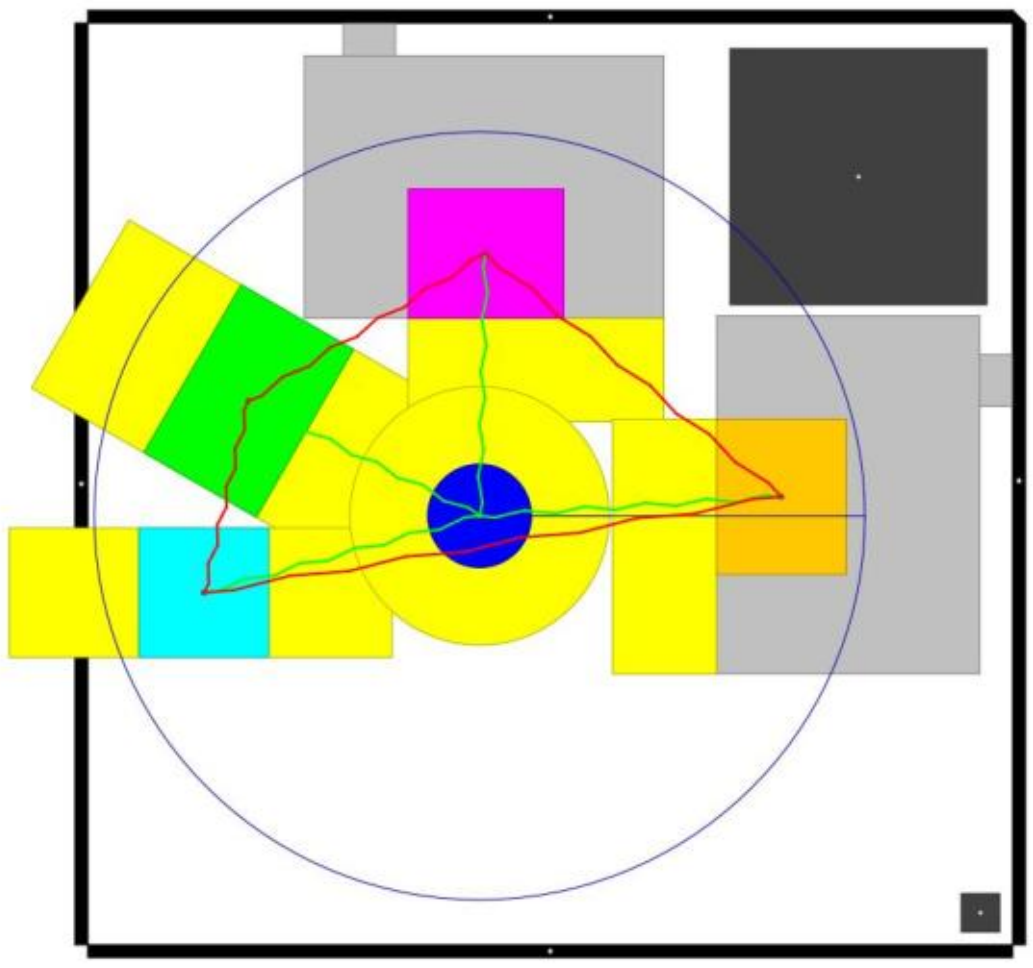

Fig. 5. Initial MSDLO layout model with "reachability" spring-damper-links rendered green and "closeness" spring-damper-links rendered red.

## 5. Results and Discussion

To account for the GA's stochastic nature, 31 experiments are carried out. It should be noted that the initial population of the GA had to be seeded with 10 valid layout solutions for it to converge and that the best configuration of the GA (see) was determined using the design for experiments.

On average each experiment required 5 min and 41 s computation time. The results of the GA are summarized in .

Table 2. Summary of the GA's results for n = 31 experiments.

| Variable | Handling Cost [€] | Area Savings [€] | Aggregated KPI [€] |
|---|---|---|---|
| Mean | -2,271.21 | +4,568.86 | +2,297.64 |
| Std. Deviation | 168.93 | 351.67 | 478.87 |
| Minimum | -2,564.56 | +3,848.22 | +1,332.09 |
| Maximum | -1,965.34 | +5,268.54 | +3,156.52 |
| 95% Confidence Interval for Mean | -2,333.17 to -2,209.26 | +4,439.88 to +4,697.84 | +2,122.05 to +2,473.27 |

Combining the two approaches (GA as a coarse optimizer and MSDLO as a subsequent local optimizer) yielded the improvements listed in Table 3.

While the average Handling Cost decreased by 11.06% and the usable area experienced minimal change (+0.17%), the Area Savings saw a decline of 15.15%. A detailed examination of the results revealed that in nine out of 31 cases, the MSDLO encountered edge cases where a usable area was divided into multiple sections due to the distance to an obstacle or boundary falling below 0.50 m. By excluding these nine edge cases (experiments #05, #08, #11, #14, #15, #16, #22, #23, and #27) from consideration, the results presented in Table 4 can be observed.

Table 3. Changes to original KPIs when applying the MSDLO to GA's Greenfield layout solutions (red-marked figures indicate a worsening of KPI).

| Variable | Reduction in Handling Cost [%] | Increase in Area Savings [%] | Change in Aggregated KPI [€] | Increase in Aggregated KPI [%] |
|---|---|---|---|---|
| Mean | -11.06 | 0.17 | -15.15 | -473.35 |
| Std. Deviation | 6.61 | 3.72 | 28.47 | 1,315.43 |
| Minimum | -0.07 | -7.94 | -70.03 | -2,747.24 |
| Maximum | -23.15 | +11.85 | 25.10 | +1,537.02 |
| 95% Confidence Interval for Mean | -8.63 to -13.48 | -1.19 to 1.54 | -25.59 to -4.72 | -919.79 to 45.09 |

Table 4. Changes to original KPIs when applying the MSDLO to GA's Greenfield layout solutions, after filtering out edge cases (red-marked figures indicate a worsening of KPI).

| Variable | Reduction in Handling Cost [%] | Increase in Area Savings [%] | Change in Aggregated KPI [€] | Increase in Aggregated KPI [%] |
|---|---|---|---|---|
| Mean | -11.38 | 2.06 | +348.84 | 20.05 |
| Std. Deviation | 7.58 | 7.29 | 448.10 | 28.81 |
| Minimum | -0.07 | -7.72 | -361.11 | -12.28 |
| Maximum | -23.15 | 25.1 | +1,537.02 | 108.85 |
| 95% Confidence Interval for Mean | -8.02 to -14.74 | -1.17 to 5.29 | +150.12 to +547.55 | 7.28 to 32.28 |

This means that for an average simulation time of 13 seconds, there is a 95% chance of improving the Aggregated KPI by at least 7.28%. If no improvement is observed, the results can be ignored, and the additional cost of -0.001 € (13 s simulation time at 0,40 €/CPUh for r6a.xlarge EC2 AWS Instance) is negligible.

## 6. Conclusions and Future Work

The results support the decision to combine the two solution approaches since the additional computation cost and time for the MSDLO method is negligible. The MSDLO method promises an increase in Aggregated KPI with a 95% confidence by at least 7.28% (= 150.12 €). That would be a return on investment of 15,012,000%. The average increase is 20.05% (= 348.48 €), i.e., improvements could be even better.

In future work, the MSDLO method will be applied to solve a Brownfield layout problem involving n= 20 resources and compared again against a GA, as well as the GA/MSDLO hybrid approach.

## Acknowledgements

The authors thank Dr. Kai Pfeifer for his insightful discussions and support. This work and its results would not have been possible without the support and guidance of Prof. T. Bauernhansl, the director of our institute. The authors used ChatGPT and Grammarly to enhance the clarity and conciseness of this paper. After using this tool/service, the authors reviewed and edited the content as needed and take full responsibility for the publication's content.